\documentclass[conference]{IEEEtran}
\ifCLASSINFOpdf
\else
\fi

\usepackage[cmex10]{amsmath}

\usepackage{amsthm}
\usepackage{amsfonts}
\usepackage{mathtools}
\usepackage{graphicx}
\usepackage{cancel}
\usepackage[]{algorithm2e}

\newtheorem{theorem}{Theorem}[section]
\newtheorem{lemma}[theorem]{Lemma}

\newtheorem{corollary}[theorem]{Corollary}
\newtheorem{example}{Example}
\usepackage{amssymb}
\usepackage{mathtools}

\DeclareMathOperator*{\plms}{p-limsup}
\DeclareMathOperator*{\plmi}{p-liminf}
\providecommand{\abs}[1]{\left\lvert#1\right\rvert} 

\begin{document}
%
\title{Secrecy through Synchronization Errors}

\author{\IEEEauthorblockN{Jason Castiglione}
\IEEEauthorblockA{University of Hawaii\\
Email: jcastig@hawaii.edu}
\and
\IEEEauthorblockN{Aleksandar Kavcic}
\IEEEauthorblockA{University of Hawaii\\
Email: kavcic@hawaii.edu}}


%


\maketitle

\begin{abstract}
In this paper, we propose a transmission scheme that achieves information theoretic security, without making assumptions on the eavesdropper's channel.  This is achieved by a transmitter that deliberately introduces synchronization errors (insertions and/or deletions) based on a shared source of randomness. The intended receiver, having access to the same shared source of randomness as the transmitter, can resynchronize the received sequence. On the other hand, the eavesdropper's channel remains a synchronization error channel. We prove a secrecy capacity theorem, provide a lower bound on the secrecy capacity, and propose numerical methods to evaluate it.
\end{abstract}


%
\IEEEpeerreviewmaketitle

\section{Introduction}
With the appearance of wireless communications in the smallest of devices, communicating securely has become of paramount interest. We see an abundance of small devices with limited computing capability having access to potentially sensitive data. In addition to limited power and computational capability, another limitation that these devices have in common is that they transmit over noisy channels. They all require error correction codes and specialized processors for reliable communications. A natural question arises, can we combine error correction codes and security?  \par
 Wyner's seminal paper \cite{wtap:wyner} proved it was possible to communicate securely using wiretap codes, assuming the composite channels were discrete and memoryless. Csiszar and Korner \cite{wtap:krner} generalized Wyner's results to broadcast channels, where the unintended user does not neccesarily listen to the output of the main channel through a noisier channel. Hayashi \cite{wtap:haya} further generalized the preceeding results to arbitrary channels with finite output alphabet. Hayashi's results were based upon using the information spectrum approach originating with Han and Verdu \cite{reso:hanver}. His results were particularly interesting since he included non-asymptotic results on secrecy capacity. Bloch and Laneman \cite{wtap:bllan1} built upon Csiszar and Hayashi's work in using channel resolvability as a basis for stronger secrecy results based on variational distance and hold for the most general of channels, e.g., arbitrary alphabets, memory, etc... Specifically, the system we present in this paper has memory, and a countably infinite output alphabet.

\begin{figure}[!htbp]
\centering
\includegraphics{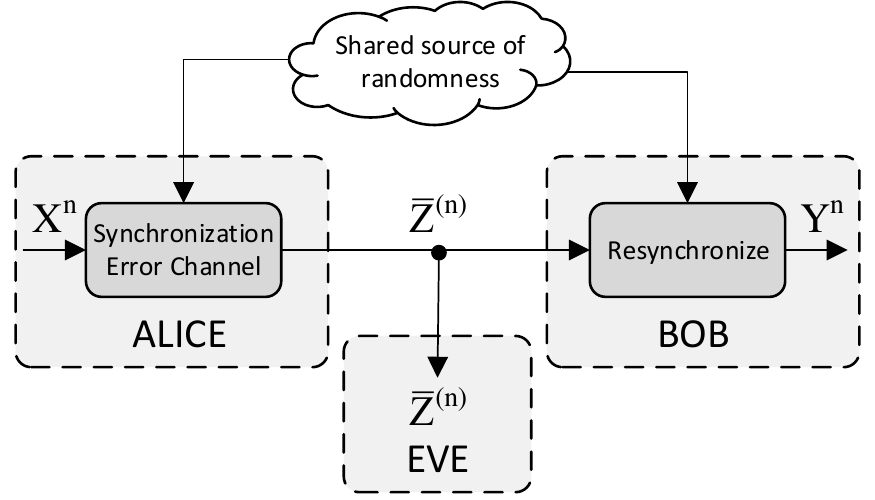}
\caption{Synchronization Error Based Secrecy System}
\label{psol}
\end{figure}

The main obstacle to achieving secrecy through wiretap codes over wiretap (or similar) channels \cite{wtap:wyner}, \cite{wtap:krner}, \cite{wtap:bllan1} is that the secrecy proof hinges on \emph{knowing, and guaranteeing the eavesdropper's channel characteristics}. In practice, the eavesdropper is typically an adversary and will not reveal her channel characteristics. For this reason, we employ a strategy in which we assume that both the intended user and the eavesdropper receive (noise-free)\footnote{In practice, this can be guaranteed, even over noisy channels, using error-correcting codes to achieve error-free transmission of information.} the same symbols transmitted by the transmitter (see Fig. 1).  Instead of transmitting the codeword $X^n$, the transmitter uses a secret shared source of randomness to randomly inject synchronization errors into the transmitted sequence. The intended receiver has access to the shared source of randomness and can thus resynchronize the received sequence, although not necessarily recover $X^n$. The eavesdropper, on the other hand, does not have access to the shared source of randomness\footnote{ For future considerations, we would like to replace the secret shared source of randomness with finite length keys drawn from a finite key space.}, and is thus forced to attempt to decode the transmitted codeword. Under this constructed scenario, as we show in this paper, the Bloch-Laneman secrecy results \cite{wtap:bllan1} hold, and we therefore can communicate securely.   \par

This method is similar to the one in \cite{misa}, but instead of injecting additive errors, the transmitter injects synchronization errors, which has the following advantageous features. \vspace{-0.1in}
\begin{itemize}
\item Synchronization error channels are cryptographically favorable \footnote{ Knowledge of the plaintext, does not guarantee an exact solution for the corresponding key. } to additive noise channels. \vspace{-0.06in}
\item We do not make unverifiable physical channel assumptions, for either the main or eavesdropper channel.\vspace{-0.06in}
\item Inclusion of wiretap codes has low overhead because systems already use error correction codes. \vspace{-0.06in}
\item Assuming a secret shared source of randomness, we are guaranteed a stronger version of secrecy then Wyner \cite{wtap:wyner} based on variational distance. \vspace{-0.06in}
\end{itemize}
Of course, these advantages must be traded off with some undesirable properties, such as, \vspace{-0.06in}
\begin{itemize}
\item We communicate at a different rate compared to conventional cryptographic systems, where the length of the plaintext and ciphertext are typically equal.\vspace{-0.06in}
\item We must construct resolvability codes for channels with synchronization errors. \vspace{-0.06in}
\item In practice, we will require a secret preshared key in place of a secret shared source of randomness.\vspace{-0.06in}
\end{itemize}

\subsection{Outline}
In Section II, we present our transmission scheme, and an equivalent wiretap channel based on synchronization errors with insertions and deletions. We then apply  Bloch and Laneman \cite{wtap:bllan1} to lower bound the secrecy capacity in Section III. \par
 In Section IV, we present hidden Markov techniques that allow us to estimate and bound the corresponding information rates. Due to the inherent difficulty of computing such rates in closed form for synchronization error channels, e.g., insertion/deletion channels, in this paper we also reveal a reduced state technique to lower bound the secrecy capacity. The lower bounding technique is constructive in the sense that it reveals the source distribution that achieves the bound. Finally, in Section V, we plot lower bounds for the secrecy capacity of the erasure/deletion, and insertion wiretap channels.
\subsection{Notation}
 An alphabet of symbols is denoted by a calligraphic letter, say $\mathcal{X}$. Given two elements $a,b \in \mathcal{X}$, we denote their \textit{concatenation} by $a \| b \triangleq ab$. We denote the length zero word, by $\theta$, where $a \| \theta = \theta \| a =a $. An alphabet $\mathcal{X}$ does not necessarily  contain $\theta$. An $n$-fold Cartesian product of $\mathcal{X}$ is denoted by $\mathcal{X}^n$. The alphabet obtained by the $n$-fold concatenation of symbols from $\mathcal{X}$ is denoted by $\mathcal{X}^{(n)}$, and by definition $\mathcal{X}^{(0)}= \{ \theta \}$. Observe if $\mathcal{X}=\{ a ,b \}$, then $\mathcal{X}^{(2)}= \{ aa,ab,ba,bb \}$. Given a base alphabet, $ \mathcal{X}$, and $x_1,x_2, \dots, x_n \in \mathcal{X}$, we define the \textit{length} of $y= x_1 \| x_2 \| \dots \| x_n=x_1  x_2 \dots x_n$, as 
\begin{equation}
L(y)= \min\{ k | k \in \{ 0 ,1 ,\dots \} \text{ such that } y \in \mathcal{X}^{(k)} \}
\end{equation}
Given an alphabet $\mathcal{X}$, we form $\mathcal{\overline{X}}$ as the alphabet that consists of all finite-length concatenations of symbols from $\mathcal{X}$, i.e., $\mathcal{\overline{X}}= \bigcup\limits_{i=0}^{\infty} \mathcal{X}^{(i)}$. Note, by construction $\mathcal{\overline{X}}^{(n)}=\mathcal{\overline{X}}$, for $n \geq 1$.

Random variables are denoted by upper-case letters, $(X)$, and their realizations by lower case letters $(x)$. If $k$ denotes discrete time, then $X_k$ denotes a random variable at time $k$ drawn from the alphabet $\mathcal{X}$. Likewise, $\overline{X}_k$ is assumed to be drawn from $\mathcal{\overline{X}}$. A sequence of random variables $X_1, X_2, \dots, X_n$ is denoted by $X^n$, and we have $X^n \in \mathcal{X}^n$. A concatenation of symbols $X_1 \| X_2 \| \dots \| X_n =X_1 X_2 \dots X_n$ is denoted by $X^{(n)}$, and we have $X^{(n)} \in \mathcal{X}^{(n)}$. Given two random variables, $X$ and $Y$, the \emph{information-spectrum} is a random variable, denoted by $I(X;Y)$. The expected value of $I(X;Y)$ will be denoted as $\mathbb{I}(X;Y)=\mathbb{E}[I(X;Y)]$. Similarly, $H(X)$ , a random variable, denotes the \emph{entropy-spectrum}, and $\mathbb{H}(X)$ denotes the expected value. We denote the binary entropy function as follows\footnote{Without loss of generality, we will assume all logarithms are base 2.}, $h(x) \triangleq -x\log(x) - (1-x)\log(1-x)$.

\section{Transmission Wiretap Model}

\subsection{Transmission Scheme}

Let $X_k$ be the channel input drawn from the finite alphabet, $\mathcal{X}= \{0,1\}$ (see Fig. 1). Our synchronization error channel has insertions and deletions with respective probabilities $i$ and $d$. Let $\overline{Z}_k$ be the transmitter output drawn from the alphabet $\mathcal{\overline{Z}}$, where $\mathcal{Z}=\mathcal{X}$. The symbols $\overline{Z}_k$ are drawn using the shared source of randomness, $\mathbf{R}$, as in algorithm 1.
\begin{algorithm}[]
\SetAlgoLined
\KwData{ $\mathbf{R}, X_1, X_2, \dots $ }
\KwResult{Transmit $\overline{Z}_1, \overline{Z}_2 \dots$  }
\While{$X_k$ at input}{
using $\mathbf{R}$ generate a geometric random variable $N_k \in \{0, 1, \dots\} $ with probability 
$(1-i)$\;
generate $N_k$ Bernoulli($\frac{1}{2}$)random variables, $B_1, B_2, \dots , B_{N_k}$ \; 
using $\mathbf{R}$ generate a Bernoulli($d$) random variable $D_k$\;
\eIf{$D_k == 1$}{
$\overline{Z}_k= B_1 \| B_2 \| \dots \| B_{N_k}$\;
}{
$\overline{Z}_k= X_k \| B_1 \| B_2 \| \dots \| B_{N_k}$\;
}
Transmit $\overline{Z}_k$
}
\caption{Transmission Scheme}
\end{algorithm}

 Since the intended receiver has access to the shared source of randomness, it can generate $N_1, N_2, \dots$ , and $D_1, D_2, \dots$ Therefore it can  create a sequence, $Y_1,Y_2,\dots$ , by removing inserted symbols and substituting all deletions by erasures (denoted by $\varepsilon$), i.e, the alphabet for $Y_k$ is $\mathcal{Y}=\{ \varepsilon \} \cup \mathcal{X}$.
\begin{lemma} \label{lemm}  $ \mathbb{E} \left[L \left( \overline{Z}^{(n)} \right) \right] =n \cdot \frac{1 - (1-i)d}{(1-i)}$
\end{lemma}
\begin{example}\textbf{Insertions and Deletions}
Let $\overline{Z}^{(n)}$ denote the sequence of symbols $\overline{z}$ generated by the transmitter after processing the first $n$ input symbols $x_1,x_2, \dots , x_n$. For example, if $(x_1,x_2,x_3, x_4)=(0,1,1,0)$ and the transmitter decides to  a) transmit $x_1$ and insert symbols $1,1$, b) transmit $x_2$ without any insertions, c) delete $x_3$, and d) transmit $x_4$ and insert a symbol $0$ (see Fig. 3). We have that $\overline{z}^{(1)}=011$, $\overline{z}^{(2)}=0111$, $\overline{z}^{(3)}=0111$, and $\overline{z}^{(4)}=011100$. Thus Alice transmits $011100$. After resynchronization, Bob has $(y_1,y_2,y_3,y_4)=(0,1, \varepsilon ,0)$.
\end{example}
\begin{figure}[!hbp]
\begin{center}
\begin{tabular}{ rlcc|c|c|c|c| } 
$x^4=$&$(0,1,1,0)$ & & $x_1$ & $x_2$ & $x_3$ & $x_4$\\ 
$y^4=$&$(0,1,\varepsilon,0)$ & &$0$ & $1$ & $1$ & $0$ \\
$\overline{z}^{(4)}=$ & $011100$  & & $ 0 \underline{11} $  & $1 $ & $\cancel{1} $ & $0 \underline{0} $\\  
$z_1,z_2,\dots,z_6=$&$0,1,1,1,0,0$& & $011$ & $1$ & & $00$ \\
\end{tabular}
\end{center}
\caption{Example 1}
\label{ptab}
\end{figure}
\subsection{Equivalent Wiretap Channel Model}
An equivalent wiretap channel model, in the terms of Bloch and Laneman \cite{wtap:bllan1}, consists of the main channel being an erasure channel, and the wiretap channel being an insertion channel followed by an $\varepsilon$-deletion channel. We define an $\varepsilon$-deletion channel as a channel that deletes the erasure symbols, $\varepsilon$. In particular, note a deletion channel is equivalent to an erasure channel followed by an $\varepsilon$-deletion channel.

\begin{figure}[!tbp]
\centering
\includegraphics{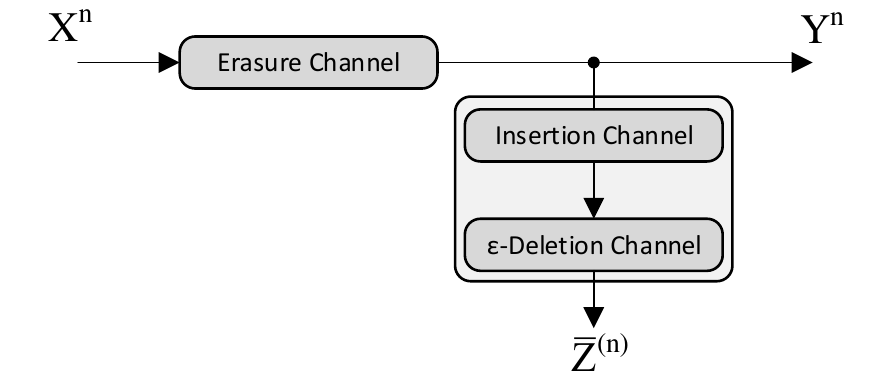}
\caption{Equivalent Degraded Wiretap  Insertion/Deletion Model}
\label{psol3}
\end{figure}

\section{Secrecy Capacity}
A main strength of Bloch's and Laneman's proof \cite{wtap:bllan1} of secrecy using information spectra, and channel resolvability is the generality of their proof. In particular their notation uses $\overline{Z}^n$ for channel output, i.e., the output is synchronized with the channel input. Similarly to Han \cite{reso:han} p. 100, we claim that the results \cite{wtap:bllan1} apply to more general channels, e.g., the synchronization error wiretap model with output $\overline{Z}^{(n)}$. 
\begin{theorem}{(Bloch-Laneman Secrecy Capacity)\cite{wtap:bllan1}} The secrecy capacity, $C_s$, of the synchronization error wiretap model for secrecy metric $\mathbb{S}_2$  is,
\begin{align}
 C_s =  \sup_{ \{V^n,X^n\}_{n \geq 1} } & \left[ \plmi_{n \rightarrow \infty} \frac{1}{n}  I(V^n;Y^n) \right.  \nonumber \\ 
& \left. -  \plms_{n \rightarrow \infty} \frac{1}{n} I(V^n;\overline{Z}^{(n)}) \right] \label{eq1} 
\end{align}
where the process $\{V^n, X^n\}_{n \geq 1}$ satisfies,
\begin{align*}
V^n \rightarrow X^n  \rightarrow \overline{Z}^{(n)} Y^n \quad \forall n \in \mathbb{Z}^{+}. \nonumber
\end{align*}

\begin{proof}
The proof in \cite{wtap:bllan1} is written for channels in which $\overline{Z}^{(n)}=\overline{Z}^n$, and uses two corollaries in Pinsker \cite{pinsk}. Note that the two Pinsker\cite{pinsk} corollaries hold for synchronization error channels if we use $\overline{Z}^{(n)}$ instead of $\overline{Z}^n$. Similarly, Bloch and Laneman's proof that, $3 \succeq 4 $, for secrecy metrics holds. The remainder of their proof relies on results in channel resolvability from Han \cite{reso:han} which hold for channels with synchronization errors.
\end{proof}
\end{theorem}
\begin{corollary}{(Lower Bound for the Secrecy Capacity of the Insertion/Deletion Wiretap Channel Model)} Let $\mathcal{M}$ represent the set of all homogeneous Markov chains defined on the alphabet $\mathcal{X}$. Then
\begin{align}
C_s &\geq \max_{ \{ X^n \} \in \mathcal{M}} \lim_{n \rightarrow \infty} \frac{1}{n} \left[  \mathbb{I}(X^n;Y^{(n)}) -    \mathbb{I}(X^n;\overline{Z}^{(n)}) \right] \label{eq2}   
\end{align}

\begin{proof}
In our channel model we have assumed the wiretap channel is a degraded version of the main channel. Thus the supremum \eqref{eq1} is attained for $V^n=X^n$. \par

If $X^n$ is a homogeneous Markov process, the limit $I=\lim_{n \rightarrow \infty} \frac{1}{n} \mathbb{I}(X^n; \overline{Z}^{(n)})$ exists and is finite (see general proof in \cite{del:dobr}, but without $\sup$ in equation 2.7). Furthermore, as in \cite{del:dobr}, p 17-19, the two sequences, $X^n$ and $\overline{Z}^{(n)}$, satisfy,
\begin{equation}
\lim_{n \rightarrow \infty} P \left(  \abs{ \frac{I(X^n;\overline{Z}^{(n)})}{nI} - 1  } > \delta \right)=0 \qquad ,\forall \delta > 0
\end{equation} 
This is obtained in a similar fashion as in Dobrushin \cite{del:dobr}, but again without the $\sup$ in eq. 2.7 of \cite{del:dobr}. It follows that,
\begin{align}
\plms_{n \rightarrow \infty} \frac{1}{n} I(X^n;\overline{Z}^{(n)}) &=\lim_{n \rightarrow \infty} \frac{1}{n} \mathbb{I}(X^n; \overline{Z}^{(n)})=I \nonumber
\end{align}
Observe the main channel is a discrete memoryless channel. Furthermore, given a stationary and ergodic input process, the output process is clearly stationary and ergodic. Thus,
\begin{equation}
\plmi_{n \rightarrow \infty}\frac{1}{n} I(X^n;Y^n) =\lim_{n \rightarrow \infty} \frac{1}{n} \mathbb{I}(X^n;Y^n).
\end{equation}
Finally, the inequality in \eqref{eq2} follows because we restricted the input to homogeneous Markov processes.
\end{proof}
\end{corollary}
 
Our intent in looking at the combined insertion/deletion channel, is to balance the output. We would like the ability to satisfy power requirements of the transmitter or receiver by altering the deletion and insertion probabilities. We shall now look at two contrasting cases, namely $i=0$ and $d=0$.
 
\subsubsection{Insertion Wiretap Channel}
The $d=0$ scenario is attractive if the receiver has power (and computational) restrictions. We see that once in sync with the transmitter, the receiver is only required to demodulate bits declared as non-insertion bits by the shared random source, i.e., $X^n=Y^n$. Thus the decoder is easy to implement on a low power device. The secrecy capacity, which we denote with $C_{si}$, is bounded as
\begin{align}
C_{si} & \geq  \lim_{n \rightarrow \infty} \frac{1}{n} \left[  \mathbb{I}(X^n;Y^n) -  \mathbb{I}(X^n;\overline{Z}^{(n)})\right] \\
& = \lim_{n \rightarrow \infty} \frac{1}{n} \left[ \mathbb{H}(X^n) - \mathbb{H}(\overline{Z}^{(n)}) + \mathbb{H}(\overline{Z}^{(n)}|X^n) \right],
\end{align}
where $X^n$ is an arbitrary Markov process.

\subsubsection{Deletion/Erasure Wiretap Channel}
For the deletion/erasure wiretap model we have $i=0$. This is the dual case to the insertion wiretap channel in the sense that the deletion of bits is relatively easy to implement at the transmitter. The main channel is now a discrete memoryless erasure channel for which there are well known codes, yet the receiver still requires some computational power to decode. This is a good choice if the transmitter is power limited (because it transmits fewer bits) but the receiver has ample power/computational resources. The secrecy capacity for this model, denoted by $C_{sd}$, is now bounded as
\begin{align}
C_{sd} & \geq \lim_{n \rightarrow \infty} \frac{1}{n} \left[  \mathbb{I}(X^n;Y^n) -  \mathbb{I}(X^n;\overline{Z}^{(n)})\right] \\
&= \lim_{n \rightarrow \infty} \frac{1}{n} \left[ \mathbb{H}(Y^n) -n\cdot h(d) -  \mathbb{H}(\overline{Z}^{(n)}) + \mathbb{H}(\overline{Z}^{(n)}|X^n) \right], \nonumber
\end{align}
where $X^n$ is an arbitrary Markov process. 

\section{Numerically Bounding $C_s$}

In this section we construct numerical techniques to lower bound $C_{si}$ and $C_{sd}$ of the insertion and deletion eavesdropper's channels given in the previous section, respectively. We shall assume the input process $X^n$ is an $M$-th order binary  Markov process, i.e., $\mathcal{X}= \{0,1\}$, whose $2^M \times 2^M$ transition probability matrix is $\mathbf{P}$. The output of the eavesdropper's channel is $\overline{Z}^{(n)} \in \mathcal{\overline{Z}}=\mathcal{\overline{X}}$. The main channel ($X^n \rightarrow Y^n$) is a memoryless erasure channel, and $Y^n \in \mathcal{Y}^n= \{ \varepsilon , 0, 1\}^n$.
\subsection{Computing $\lim \frac{1}{n} \mathbb{H}(X^n)$}
Being Markov, $X^n$ has a closed form entropy rate \cite{inft:cover}. 
\subsection{Computing $\lim \frac{1}{n} \mathbb{H}(Y^n)$}
Since $X^n$ is a Markov Process, $Y^n$ is a hidden Markov process obtained by passing $X^n$ through an erasure channel with erasure probability $d$. The entropy rate of $Y^n$ can be evaluated using trellis-based Monte Carlo techniques \cite{est:kklvz}.
 
 \subsection{Computing $\lim \frac{1}{n} \mathbb{H} \left( \overline{Z}^{(n)} \right) $}

Using Lemma \ref{lemm}, we have 
\begin{equation}
\lim_{n \rightarrow \infty} \frac{1}{n} \mathbb{H} (\overline{Z}^{(n)})= \frac{1-d+di}{1-i} \lim_{k \rightarrow \infty} \frac{1}{k} \mathbb{H}(Z^k)
\end{equation}
where $Z^k$ is the sequence of symbols $Z_k$ at the output of the eavesdropper's channel. [Note, $Z^k\in \mathcal{Z}^k=\{0,1\}^k \text{, whereas } \overline{Z}^{(n)}\in \mathcal{\overline{Z}}$.] Next, we notice that the sequence $Z^k$ is itself a hidden Markov process with an underlying Markov state sequence $S^k$ whose transition probability matrix is $\mathbf{Q}$ (explicitly given further down). That is,
\begin{equation}
Pr(z^k | s_0 ,s^k)= \prod_{m=1}^k Pr(z_m| s_{m-1},s_m)Q(s_{m-1},s_m)
\end{equation}
where $Q(s_{m-1},s_m)$ are entries in $\mathbf{Q}$. The state $s_m$ is described as a binary string. If $X^k$ is a Markov source of order M, then $s_m$ is a binary string of length $M$. We denote by $l(s_m)$ the last (i.e. the $M$-th) binary digit of $s_m$. With this notation, we can fully describe the hidden Markov process $Z^k$ as follows.
\subsubsection{Insertion channel}  We have,
\begin{equation}
\mathbf{Q}=\mathbf{Q}_i=(1-i)\mathbf{P}+i\mathbf{I} \text{ ,} \qquad  \text{and}
\end{equation}
\begin{align}
	Pr & (z_m|s_{m-1},s_m)= \\ \nonumber
	& \left\{
		\begin{array}{lll}
			\frac{i/2}{Q(s_{m-1},s_{m})} & \text{if } s_{m-1}=s_{m} \text{ and } z_m \neq l(s_m) \\
			 1- \frac{i/2}{Q(s_{m-1},s_{m})} &\text{if } s_{m-1}=s_{m} \text{ and } z_m = l(s_m) \\
			1 &  \text{for all other valid pairs } (s_{m-1},s_m)
		\end{array}
	\right.
\end{align}
\subsubsection{Deletion channel} We have,
\begin{equation}
\mathbf{Q}=\mathbf{Q}_{d}= (1-d) \left[ \mathbf{I} -d \mathbf{P}\right] ^{-1} \mathbf{P} \text{ ,} \qquad  \text{and}
\end{equation}
\begin{equation}
	Pr(z_m|s_{m-1},s_m)=\left\{
		\begin{array}{lll}
			1 &   \text{if }z_m = l(s_m) \\
			0 &  \text{if }z_m \neq l(s_m) \\
		\end{array}
	\right.
\end{equation}
Since $Z^k$ is hidden Markov, its information rate is computable using trellis-based Monte-Carlo techniques \cite{est:kklvz}.
\subsection{Bounding $\lim \frac{1}{n}\mathbb{H}\left( \overline{Z}^{(n)} |X^n \right)$}
\subsubsection{Insertion Wiretap Channel} For the insertion channel, there are finitely many possible states for each received $z_m$, i.e., $(k+1)$ possible states at time $k$ (with regards to the receiver). For this paper we did not apply reduced state techniques, which is feasible for $k\approx 10^6$.

\subsubsection{Deletion Wiretap Channel} For the deletion channel,
we do not have a method to compute the conditional entropy rate 
$ \lim\limits_{n \rightarrow \infty}\frac{1}{n}\mathbb{H}\left( \overline{Z}^{(n)} |X^n \right)$, so we resort to a reduced-state technique to lower bound the conditional entropy rate. For this purpose, we first need to formulate an appropriate trellis (with possibly countably infinite number of states per trellis section), then we apply an appropriate reduced-state technique to reduce the number of states per trellis section to a finite number and guarantee the lower bound on the conditional entropy rate.
\begin{flushleft}
\begin{figure}[!htbp]
\includegraphics[width=3.3in,height=2.8in]{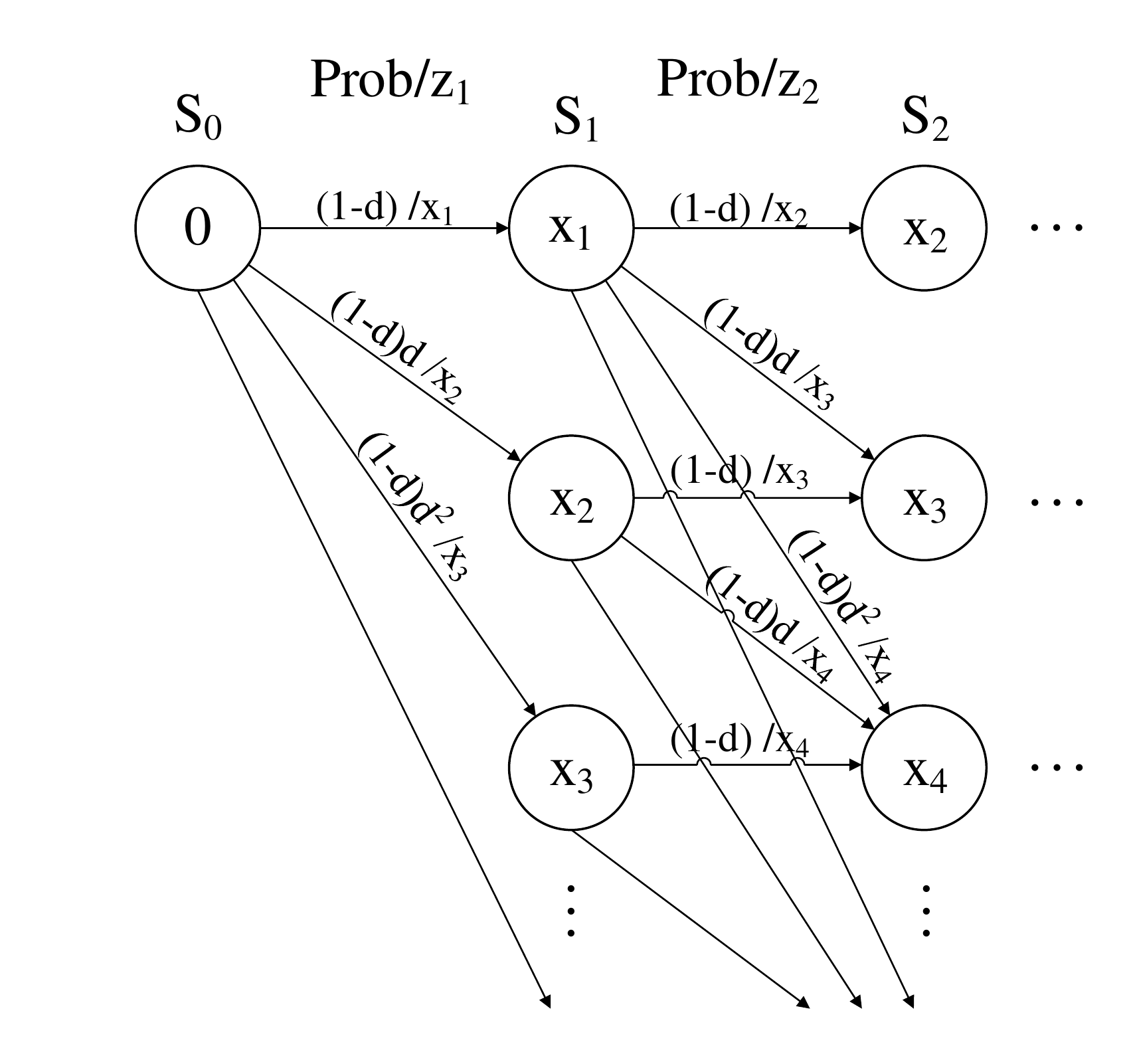}
\caption{Conditional Entropy Rate Trellis for the Deletion Channel}\label{psol4}
\end{figure}
\begin{figure}[!htbp]
\includegraphics[width=3.3in,height=2.8in]{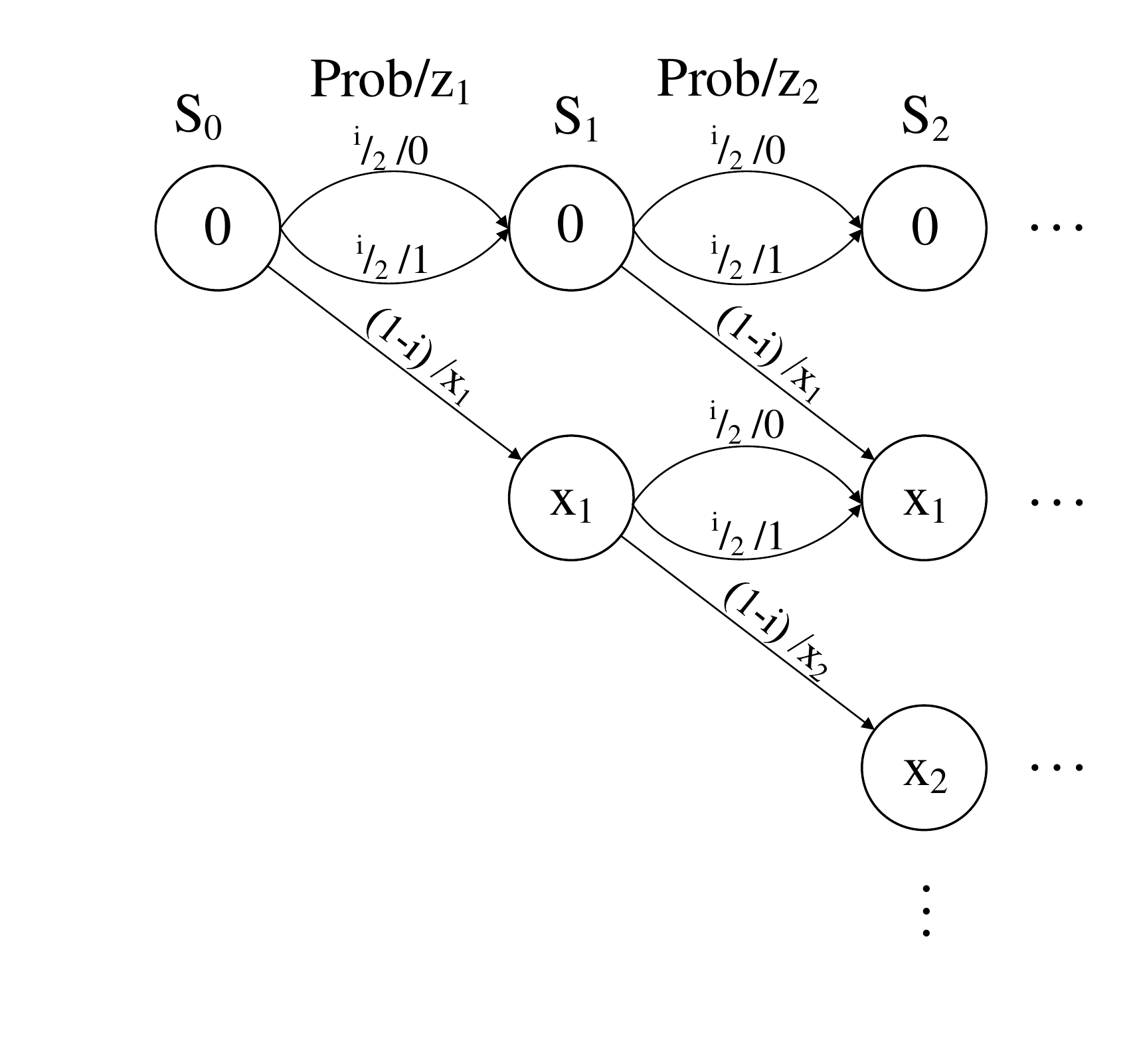}
\caption{Conditional Entropy Rate Trellis for the Insertion Channel}
\label{psol4b}
\end{figure}
\end{flushleft}
  Because of the spatial constraints, we do not fully describe the method for reducing the states, but refer the reader to \cite{est:kvmot} for general reduced-state techniques for upper bounding entropy rates.
The trellises constructed for computing the conditional entropy rate, $ \lim\limits_{n \rightarrow \infty}\frac{1}{n}\mathbb{H}\left( \overline{Z}^{(n)} |X^n \right)$, for the deletion and insertion channels have the form as in Fig.~\ref{psol4} and Fig.~\ref{psol4b}, respectively. Note that the trellises are drawn/constructed only after a realization $x_1 ,x_2, \dots$ becomes available.
\subsection{Optimization of Markov Sources}
Any Markov source will result in a lower bound for $C_{si}$ and $C_{sd}$. The best lower bound is obtained by maximizing the bound for varying Markov orders and Markov chain parameter $\mathbf{P}$. This optimization can be done by a generalized Blahut-Arimoto algorithm \cite{kavma} \cite{oncap} adapted to wiretap channels \cite{fswba}. 
\section{Numerical Results}
\begin{figure}[!htbp]
\includegraphics[width=3.5in]{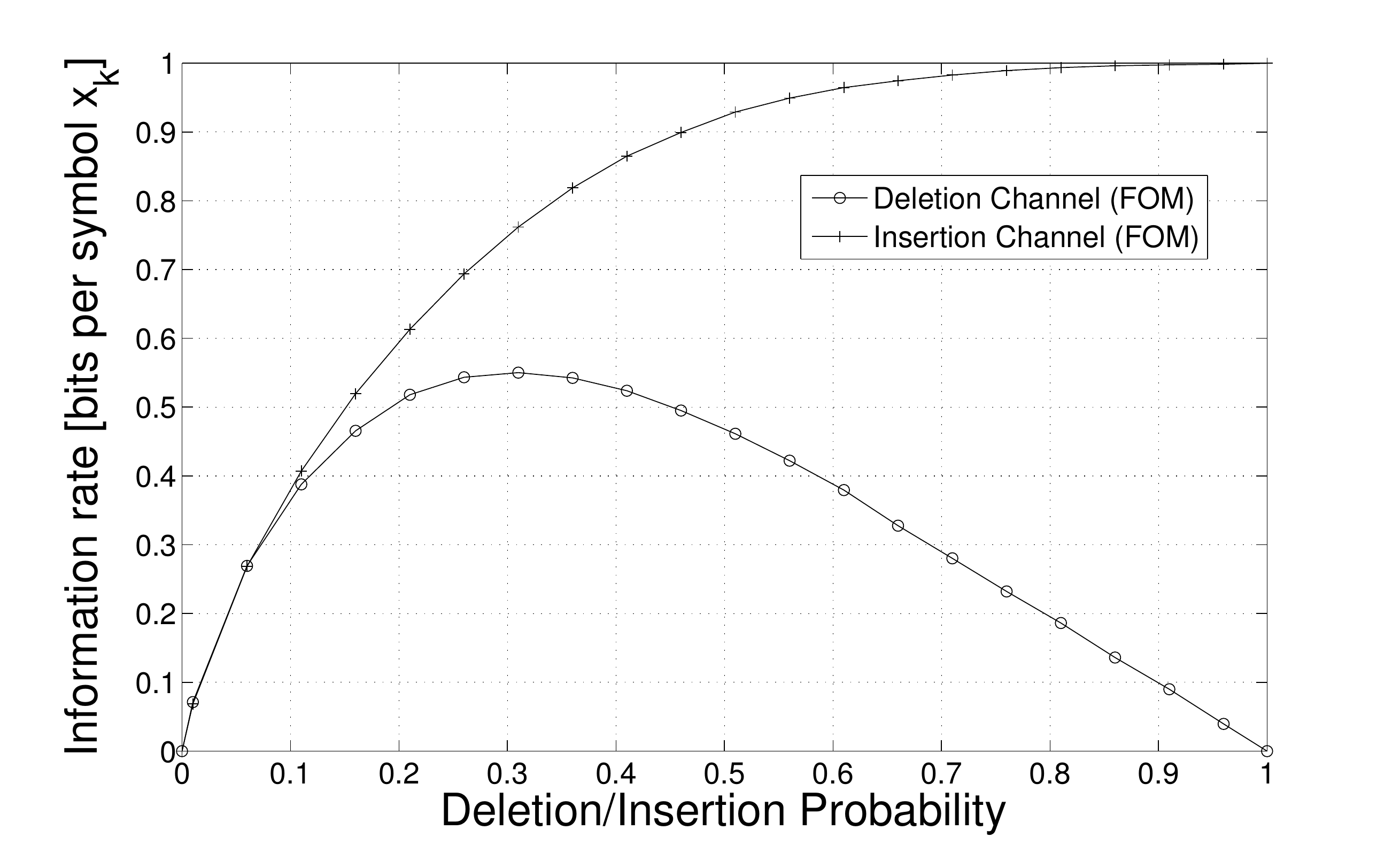}
\caption{Secrecy Capacity (Lower Bound) vs Deletion/Insertion Probability for first order Markov (FOM) inputs}
\label{psol5}
\end{figure}

In this paper, we present only numerical results (see Fig. \ref{psol5}) for Markov processes of order $M=1$, thereby we do not resort to Blahut-Arimoto-type optimizations because an exhaustive search optimization is feasible for low-order Markov processes. For both  the insertion and deletion channel we averaged $100$ simulations that were run on a trellis of length $10^5$.  \par
 We see that although $ \lim_{i \rightarrow 1}C_{si} = 1$, in our transmission scheme this is of little help since as $i \rightarrow 1$ , we are unable to transmit to the intended user. Thus we must also look at the \emph{effective transmission rate}, $R_E$, which we define as,
\begin{equation}
R_E= \frac{n}{\mathbb{E} \left[L \left( \overline{Z}^{(n)} \right) \right]}
\end{equation}
Observe, for our system we have $R_E = \frac{(1-i)}{1 - (1-i)d}$, and for $i=1$ and $d=0$, $R_E=0$. Thus even though with regards to our equivalent wiretap channel model, we have $C_{si}=1$, our implementation has $R_E=0$, and we can therefore not communicate with the intended user.
 
\section{Conclusion}
We have presented a secrecy system based on synchronization errors, and provided techniques to lower bound the secrecy capacity. Furthermore, we have evaluated these bounds (using first order Markov input) for two important instantiations of the model. The proposed system is a complimentary example to the \emph{one time pad} \cite{fmotp} \cite{wtap:shan}, in the sense that it is a method to use a shared source of randomness and achieve information theoretic security, yet it is not a stream cipher. The method has several advantages over a one time pad, e.g., the shared source of randomness is not uniquely determined given ciphertext and plaintext. \par
The transmission scheme also has advantages with regards to system design. We have the ability to choose insertion and deletion probabilities to satisfy receiver, and transmitter requirements. In particular, we see in a system with powerful base stations, and lightweight distributed sensors, the base stations can transmit based on $d=0<i<1$ and the lightweight sensors can transmit based on $i=0<d<1$. Thus the lightweight sensors transmit less, and decrypt easier, while the base stations can transmit more, and decode using computationally intensive error correction decoders.



%

\end{document}